# Puzzling properties of the historical growth rate of income per capita explained

Ron W Nielsen[1]

Griffith University, Environmental Futures Research Institute, Gold Coast Campus, Australia

**Abstract.** Galor discovered many mysteries of the growth process. He lists them in his Unified Growth Theory and wonders how they can be explained. Close inspection of his mysteries reveals that they are of his own creation. They do not exist. He created them by his habitually distorted presentation of data. One of his self-created mysteries is the mystery of the alleged sudden spurt in the growth rate of income per capita. This sudden spurt never happened. In order to understand the growth rate of income per capita, its mathematical properties are now explored and explained. The explanation is illustrated using the historical world economic growth. Galor also wonders about the sudden spurt in the growth rate of population. We show that this sudden spurt was also created by the distorted presentation of data. The mechanism of the historical economic growth and of the growth of human population is yet to be explained but it would be unproductive to try to explain the non-existing and self-created mysteries of the growth process described in the scientifically unacceptable Unified Growth Theory. However, the problem is much deeper than just the examination of this theory. Demographic Growth Theory is based on the incorrect but deeply entrenched postulates developed by accretion over many years and now generally accepted in the economic and demographic research, postulates revolving around the concept of Malthusian stagnation and around a transition from stagnation to growth. The study presented here and earlier similar publications show that these postulates need to be replaced by interpretations based on the mathematical analysis of data and on the correct understanding of hyperbolic distributions.

## Introduction

In the subsection entitled "Mysteries of the growth process" (Galor, 2005a, p. 220) presented in his Unified Growth Theory (Galor, 2005a, 2011), Galor asks a series of questions about the mysteries of economic growth. We can take his questions one by one and show that all these mysteries were of his own creation.

His theory is *not* based on the scientific analysis of data but on impressions supported by the habitually distorted presentation of data (Ashraf, 2009; Galor, 2005a, 2005b, 2007, 2008a, 2008b, 2008c, 2010, 2011, 2012a, 2012b, 2012c;

---

Galor and Moav, 2002; Snowdon & Galor, 2008). Such approach to research can easily create many mysteries that simply do not exist.

One of Galor's questions about the alleged mysteries of growth process is "What is the origin of the sudden spurt in growth rates of output per capita and population that occurred in the course of the take-off from stagnation to growth?" (Galor, 2005a, p. 220). In just one sentence, Galor presents two incorrect doctrines: the doctrine of the presence of the sudden spurt and the doctrine of the transition from stagnation to growth, both created by the failure to follow scientific principles of investigations, which require that data should *not* be manipulated to support preconceived ideas but that they should be methodically analysed with the aim of learning from them. We shall show that this question makes as much sense as the question, "Why is the Sun rotating around the Earth," and the answer to both of them is similar: the Sun does not rotate around the Earth and there was no sudden spurt in the growth rates of output per capita and population. There was also no takeoff from stagnation to growth (Nielsen, 2014, 2015, 2016a, 2016b, 2016c, 2016d, 2016e, 2016f)

We have already demonstrated that the growth of human population and the growth of the Gross Domestic Product (GDP), global and regional, were hyperbolic (Nielsen, 2014, 2016a, 2016b, 2016c, 2016d, 2016e). Hyperbolic growth is monotonic, and consequently it is also characterised by the monotonically-increasing growth rate. There is no sudden spurt in this type of distributions.

The output per capita (also described as income per capita and measured using the GDP/cap) is represented by the ratio of two, monotonically-increasing, hyperbolic distributions (Nielsen, 2015). The growth rate of this ratio is monotonic. It cannot contain "the sudden spurt" claimed erroneously by Galor.

Galor's questions about the mysteries of growth are strongly misleading because they describe features created by the distorted presentation of data. The created features and the associated questions divert attention from the correct understanding of the mechanism of economic growth. Galor's theory does not explain the mechanism of economic growth but describes phantom features he created.

We have already discussed (Nielsen, 2014, 2015, 2016a, 2016b, 2016c, 5016d, 2016e) various aspects of Galor's theory (Galor, 2005a, 2011). We shall now focus our attention on the discussion of his unsubstantiated claims about the growth rates. We shall use precisely the same data (Maddison, 2001) as used by Galor (2005a, 2011), who unfortunately did not analyse them.

Unified Growth Theory is fundamentally incorrect but it is just an embodiment of the incorrect concepts used traditionally in the economic and demographic research, concepts developed by accretion over many years and now so strongly entrenched that it will be difficult to uproot them and replace them by correct interpretations. However, it is expected that it is in the interest of every economist and demographer to have scientific basis for their research.

These erroneous interpretations revolve around the concept of Malthusian stagnation and around a transition from stagnation to growth. The study presented here and in earlier publications (Nielsen, 2014, 2015, 2016a, 2016b, 2016c, 2016d, 2016e, 2016f) demonstrate that these traditional interpretations need to be replaced by interpretations based on the mathematical analysis of data and on the correct understanding of hyperbolic distributions.

The latest data of Maddison (2001, 2010) serve as a rich source of information. When mathematically analysed, conclusions based on these data are in perfect agreement with earlier research (e.g. Biraben, 1980; Clark,1968; Cook,1960; Durand, 1967, 1974, 1977; Gallant, 1990; Haub, 1995; Kremer, 1993; Kapitza, 2006; Livi-Bacci, 1997; McEvedy & Jones, 1978; Podlazov, 2002, Shklovskii, 1962, 2002; Taeuber & Taeuber, 1949; Thomlinson, 1975; Trager, 1994; von Foerster, Mora and Amiot , 1960; von Hoerner, 1975). Their combined message is that the demographic and economic growth research has to be based on accepting the unambiguous and consistent evidence in data that the historical growth of human population and of economic growth were hyperbolic and that such a growth cannot be divided into two or three different regimes of growth governed by distinctly different mechanisms of growth. Hyperbolic growth is slow over a long time and fast over a short time but it is still the same growth governed by the same mechanism of growth. Hyperbolic distributions have to be interpreted as a whole and not in parts. What appears as stagnation is hyperbolic growth and what appears as takeoff or explosion is the natural continuation of the same hyperbolic growth.

## Fundamental mathematics

Growth rate $R(S)$ of a growing entity $S$ can be defined as:

$$R(S) \equiv \frac{1}{S}\frac{dS}{dt} ,  \qquad (1)$$

where $S$ can represent the GDP, the size of the population or any other growing entity.

Let us assume that we have two growing entities $S_1$ and $S_2$, and that we want to calculate the growth rate of the ratio of these two entities, i.e. the growth rate $R(S_1/S_2)$. It is easy to see that

$$R(S_1/S_2) \equiv \frac{1}{S_1/S_2}\frac{d(S_1/S_2)}{dt} = R(S_1) - R(S_2) . \qquad (2)$$

We have obtained an interesting and important equation. The growth rate of the ratio of two distributions is the difference between the growth rates of its two components. Thus, for instance, the growth rate of the GDP/cap is given by the difference between the growth rate of the GDP and the growth rate of population.

If two growing entities increase monotonically (as it is in the case of the historical economic growth and of the historical growth of population) their growth rates also increase monotonically and consequently the growth rate of their ratio, which is represented by the difference between the monotonically-increasing growth rates of each of the two components, is also increasing monotonically. It does not contain a sudden spurt.

Hyperbolic growth is described by the following simple differential equation:

$$\frac{1}{S}\frac{dS}{dt} = kS , \qquad (3)$$

where $S$ can represent the GDP or the size of the population, or indeed any other hyperbolically-increasing entity, while $k$ is a positive constant.

If we compere this differential equation with the general definition of the growth rate given by the eqn (1) we can see that the characteristic feature of hyperbolic growth is that its growth rate is directly proportional to the size of the growing entity:

$$R(S) = kS. \quad (4)$$

The growth rate of hyperbolic distributions increases hyperbolically. The time dependence of the growth rate of hyperbolic distributions creates precisely the same illusions as the time dependence of hyperbolic growth (Nielsen, 2014). The growth rate of hyperbolic distribution is slow over a long time and fast over a short time but it is a monotonically-increasing distribution, which cannot be divided into mathematically-justifiable slow and fast components because the transition from slow to fast growth occurs all the time along the entire time-range of such a distribution. The growth rate of hyperbolic growth does not contain any sudden spurt at any time.

The equation (3) can be solved easily by substitution $S = Z^{-1}$. Its solution is:

$$S = \frac{1}{(a - kt)}, \quad (5)$$

where $a$ is a constant, which can be determined empirically by comparing the calculated curve with data.

So, now, if we use the eqn (2) and if we assume that $S_1$ and $S_2$ are hyperbolic, then

$$S_1 = \frac{1}{(a_1 - k_1 t)} \quad (6)$$

and

$$S_2 = \frac{1}{(a_2 - k_2 t)}. \quad (7)$$

Consequntly, by using the eqn (4) we have

$$R(S_1) = \frac{k_1}{a_1 - k_1 t} \quad (8)$$

and

$$R(S_2) = \frac{k_2}{a_2 - k_2 t}. \quad (9)$$

If we now use these expressions in the eqn (2) we shall get

$$R(S_1/S_2) = \frac{k_1}{a_1 - k_1 t} - \frac{k_2}{a_2 - k_2 t} = \frac{\Delta}{(a_1 - k_1 t)(a_2 - k_2 t)}, \quad (10)$$

where

$$\Delta = k_1 a_2 - k_2 a_1. \quad (11)$$

The eqn (10) can be also presented as

$$R(S_1 / S_2) = \frac{1}{A_0 + A_1 t + A_2 t^2}, \quad (12)$$

where

$$A_0 = \frac{a_1 a_2}{\Delta}, \tag{13}$$

$$A_1 = -\frac{k_1 a_2 + k_2 a_1}{\Delta}, \tag{14}$$

$$A_2 = \frac{k_1 k_2}{\Delta}. \tag{15}$$

So, while the growth rate of hyperbolic distributions is described by the reciprocal of a linear function [see the eqns (8) and (9)] the growth rate of the ratio of hyperbolic distributions is described by the reciprocal of the second-order polynomial [see the eqn (12)]. We could call it the second-order hyperbolic distribution. It is a distribution, which resembles closely the first-order hyperbolic distribution (the reciprocal of the linear function) because it also increases slowly over a long time and escapes to infinity at a fixed time. However, it is a monotonically-increasing distribution.

It obviously makes no sense to claim a sudden spurt in the monotonically changing second-order polynomial and it obviously makes no sense to claim a sudden spurt in the reciprocal of the second-order polynomial. The sudden spurt can be created by distorting data, as Galor did, but then it is no longer science. Whether deliberately created or not, such distorted presentation of evidence is generally unacceptable even outside science. However, suitable distortion of evidence is sometimes used in defending doctrines accepted by faith. The distorted presentation of empirical evidence makes the Unified Growth Theory (Galor, 2005a, 2011) scientifically unacceptable.

Another way to understand that the growth rate of the ratio of two hyperbolic distributions (e.g. the growth rate of the GDP/cap distribution) cannot contain a sudden spurt is by looking at the denominator of the eqn (10), which is given by a product of two linearly decreasing functions. Multiplication of two linear distributions produces a monotonic distribution, which does not contain a sudden spurt.

Had Galor analysed the data (Maddison, 2001) he would have perhaps found that the GDP and the size of the population were increasing hyperbolically. Maybe, then, it would have been clear to him that monotonic distributions cannot be characterised by the non-monotonic sudden spurt. Such an analysis should have been prompted by the discovery made over 50 years ago that the growth of human population during the AD era was hyperbolic (von Foerster, Mora & Amiot, 1960). This vital discovery, published in the prestigious journal of *Science*, is not even mentioned in Galor's theory, maybe because it was an inconvenient discovery.

As in the case of hyperbolic distributions which can be studied easily by investigating the reciprocal values of the size of the growing entity, $1/S$ (Nielsen, 2014), an easy way to study the growth rate of the ratio of hyperbolic distributions is by using the reciprocal values of the growth rate, $1/R(S_1/S_2)$. As $1/S$ converts the confusing hyperbolic distribution to a linear function, which is then easy to understand, so also the reciprocal values, $1/R(S_1/S_2)$, convert the second-order hyperbolic distribution into an easy-to-interpret second-order polynomial. In both cases, the confusing features, which create the illusion of a slow growth over a long time followed by a sudden spurt disappear and are replaced by much simpler distributions.

Confusing features of hyperbolic distributions can be also clarified by using semilogarithmic displays. Such displays are routinely used for distributions, which vary over a large range of values. We shall use them in our present discussion.

## World economic growth

*Growth of the GDP, population and income per capita (GDP/cap)*

According to Galor, historical economic growth is characterised by takeoffs from stagnation to growth, which occurred around AD 1750 for developed regions and around AD 1900 for less-developed regions (Galor, 2008a, 2012a). He describes them as "stunning" or "remarkable" escapes from the Malthusian trap (Galor, 2005a, pp. 177, 220). Such remarkable escapes should be readily identifiable in data describing economic growth and of the growth of human population. In particular, for the data describing the world economic growth and the growth of population, we should see clearly two takeoffs, around AD 1750 and around AD 1900.

The alleged takeoff from the assumed stagnation to growth for developed countries coincides with the onset of the Industrial Revolution, AD 1760-1840 (Floud & McCloskey, 1994), which according to Galor was "the prime engine of economic growth" (Galor, 2005a, p. 212). It is, therefore, yet another reason why the takeoff for the developed countries and the associated "sudden spurt" in the growth rates should be easy to identify because the alleged prime engine should have been working most effectively in these countries.

Results of analysis of precisely the same data (Maddison, 2001) as used, but never analysed, by Galor during the formulation of the Unified Growth Theory (Galor, 2005a, 2011), are presented in Figures 1-3.

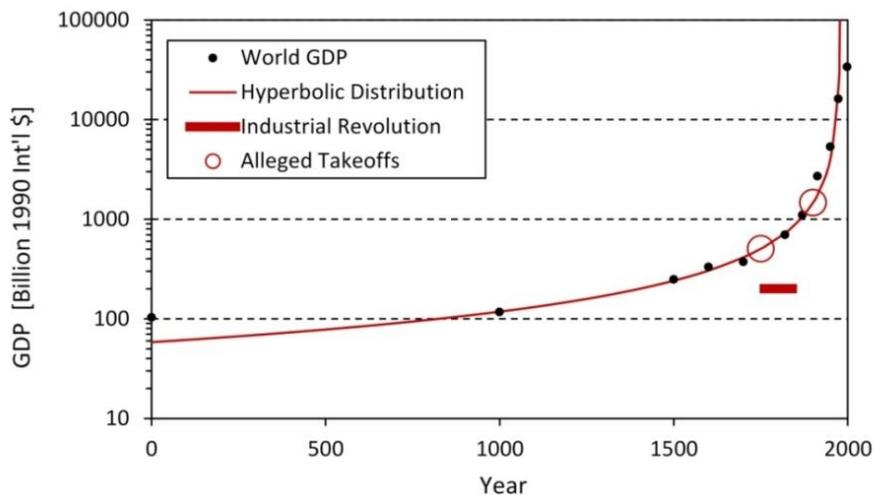

**Figure 1.** *Data for the Gross Domestic Product (Maddison, 2001), precisely the same data as used but never analysed by Galor during the formulation of the Unified Growth Theory (2005a, 2011), are compared with the first-order hyperbolic distribution [eqn (5)]. The GDP is expressed in billions of 1990 International Geary-Khamis dollars. Parameters describing the fitted hyperbolic distribution are: $a = 1.716 \times 10^{-2}$ and $k = 8.671 \times 10^{-6}$.*

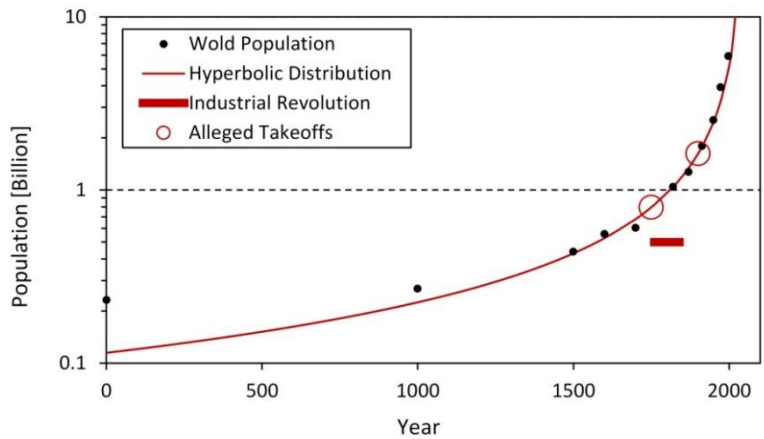

**Figure 2.** *Data describing the growth of the world population during the AD era (Maddison, 2001) are compared with the hyperbolic distribution. The large discrepancy at AD 1 is because of the maximum in the growth of the world population around that year associated with the transition from a fast-increasing hyperbolic distribution during the BC era to a slower hyperbolic distribution during the AD era (Nielsen, 2016b). Parameters describing the fitted hyperbolic distribution are $a = 8.724 \times 10^0$ and $k = 4.267 \times 10^{-3}$.*

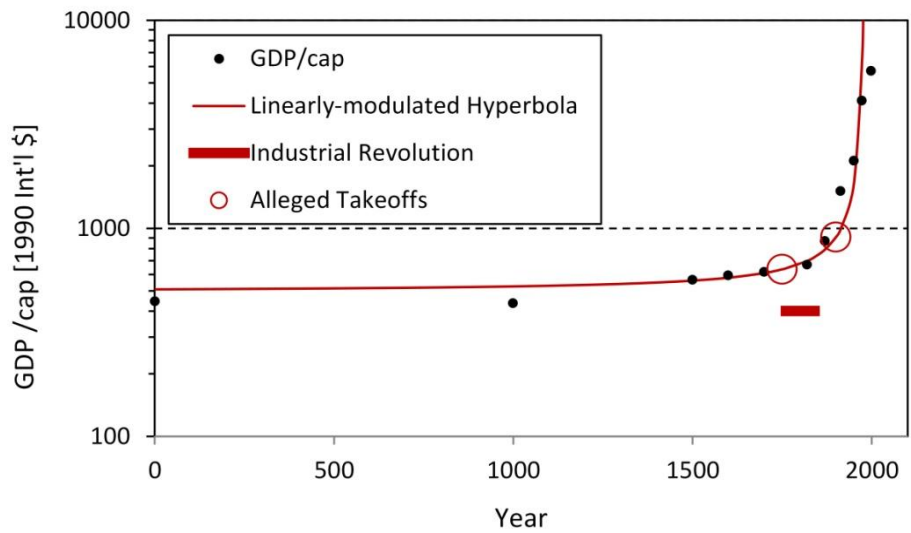

**Figure 3.** *Income per capita (GDP/cap). Data (Maddison, 2001) compared with the linearly-modulated hyperbolic distribution (Nielsen, 2015) representing the ratio of hyperbolic distributions of the GDP and of the size of population. Income per capita was increasing monotonically. Such monotonic increase cannot produce a non-monotonic growth rate claimed by Galor (2005a, 2011). It cannot produce "the sudden spurt in the growth rates of output per capita" (Galor, 2005a, p. 220). His "stunning" or "remarkable" takeoffs from stagnation to growth (Galor, 2005a, pp. 177, 220) did not happen. Industrial Revolution, the "prime engine of economic growth" (Galor, 2005a, p. 212) had no impact on changing the economic-growth trajectory. All these stories are contradicted by data (Maddison, 2001), precisely the same data as used but not analysed during the formulation of the Unified Growth Theory.*

Economic growth, as described by the Gross Domestic Product (GDP) shown in Figure 1, was hyperbolic. The alleged "prime engine of economic growth" (Galor, 2005a, p. 212) did nothing to change the growth trajectory. This is an interesting issue, which requires further investigation because technological discoveries were used to support economic growth but surprisingly perhaps they had absolutely no impact on changing the growth trajectory. It is as if economic growth was controlled by some other unknown and much stronger force, which was active before the Industrial Revolution and remained undisturbed during and after the Industrial Revolution. The alleged takeoffs from stagnation to growth did not happen because there was no stagnation. Economic growth was hyperbolic before and after the alleged takeoffs. The takeoffs claimed by Galor simply did not exist.

The growth of population during the AD era, shown in Figure 2, was also hyperbolic, at least from around AD 1000, in perfect agreement with the discovery made over 50 years ago by von Foerster, Mora and Amiot (1960). The discrepancy at AD 1 is explained by the analysis of the growth of human population in the past 12,000 years (Nielsen, 2016b), which revealed a maximum around that year associated with the transition from a fast-increasing hyperbolic growth during the BC era to a slower hyperbolic growth during the AD era. This extended analysis demonstrated that there was an uninterrupted hyperbolic growth between 10,000 BC and around 500 BC, followed by a transition to a new hyperbolic growth commencing around AD 500. It also revealed a small disturbance of the hyperbolic growth between AD 1200 and 1400. The data show that during the past 12,000 years there was no stagnation and no sudden takeoff at any time, both in the growth of the population and in the growth rate.

It is remarkable that so many independent studies are in such perfect agreement: Maddison's data (Maddison, 2001, 2010) and their analysis (Nielsen, 2016a, 2016c, 2016d, 2016e, 2016f); the estimates of the size of human population not only during the AD era but also during the BC era (e.g. Biraben, 1980; Clark,1968; Cook,1960; Durand, 1967, 1974, 1977; Gallant, 1990; Haub, 1995; Livi-Bacci, 1997; McEvedy & Jones, 1978; Taeuber & Taeuber, 1949; Thomlinson, 1975; Trager, 1994) and their analysis (e.g. Kremer, 1993; Nielsen, 2016b; Kapitza, 2006); the discovery made by von Foerster, Mora and Amiot (1960) and similar identifications of hyperbolic growth by Hoerner (1975), Podlazov (2002) and Shklovskii (1962, 2002).

In contrast, Unified Growth Theory and the generally accepted but questionable postulates used in economic and demographic research describe events and processes occurring in the world characterised by Malthusian stagnation, takeoffs, sudden spurts and by the "remarkable" or "stunning" escapes from the Malthusian traps (Galor, 2005a, pp. 177, 220), the world which is entirely different than the world revealed by data and by their mathematical analysis.

There appears to be no formal definition of Malthusian stagnation but this concept is totally irrelevant in the study of the mechanism of economic growth and of the growth of human population. They were hyperbolic. There was no stagnation and no transition from the imagined stagnation to growth. Using such descriptions to explain the mechanism of growth is unscientific because these postulates are consistently contradicted by data.

Results of analysis of income per capita (GDP/cap) presented in Figure 3 also demonstrate a monotonically-increasing distribution at least from AD 1500, i.e. during the time when Galor's "remarkable" and "stunning" effects should be clearly visible. What is remarkable about this distribution is that nothing remarkable or stunning ever happened. The growth of the GDP/cap was remarkably stable. Industrial Revolution did not accelerate the growth of income per capita and there were no sudden takeoffs at any time.

Such monotonically-increasing distributions, as presented in Figures 1-3, cannot be expected to generate "the sudden spurt" (Galor, 2005a, p. 220) in the corresponding growth rates and we shall soon see that they did not.

*Growth rates of the GDP, population and income per capita (GDP/cap)*

Results of analysis of growth rates are shown in Figures 4-6. Empirical growth rates were calculated using Maddison's data (Maddison, 2001) and interpolated gradients. The predicted growth rates were calculated using the fitted distributions shown in Figures 1-3.

As expected, the growth rate of the world GDP was increasing monotonically. Industrial Revolution did not accelerate economic growth. The "remarkable" or "stunning" escapes from the Malthusian trap (Galor, 2005a, pp. 177, 220), which were supposed to have been reflected in the takeoffs from stagnation to growth, never happened because there was no stagnation and the trap did not exist. The growth rate was increasing along a remarkably robust trajectory.

Analysis of the growth rate of the world population shows also the remarkable contradiction of Galor's claims by precisely the same data, which he used, but never analysed, during the formulation of his Unified Growth Theory. His wished-for and claimed features never happened. The growth rate of human population was increasing monotonically. There was absolutely no sudden spurt at any time.

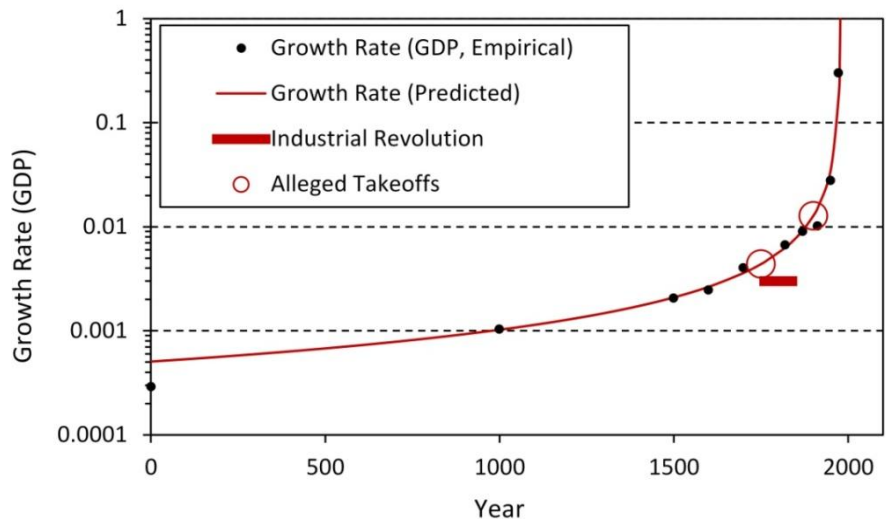

**Figure 4.** *Growth rate of the wold GDP was increasing monotonically. There was no sudden spurt. The claimed takeoffs did not happen. Industrial Revolution did not accelerate economic growth.*

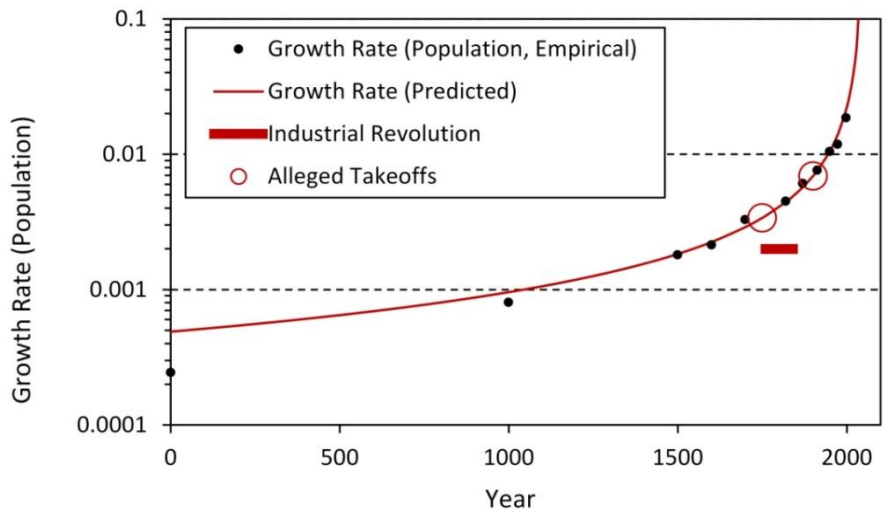

**Figure 5.** *Growth rate of the world population. Empirical growth rate calculated using Maddison's data (Maddison, 2001) and interpolated gradients is compared with the predicted growth rate calculated using parameters of the fitted hyperbolic distribution displayed in Figure 2. Galor's claims (Galor, 2005a, 2011) are remarkably contradicted by the analysis of Maddison's data (Maddison, 2001), precisely the same data, which he used but never properly analysed. Galor's mystery of "the sudden spurt" in the growth rate of population (Galor, 2005a, p. 220) is solved – there was no sudden spurt.*

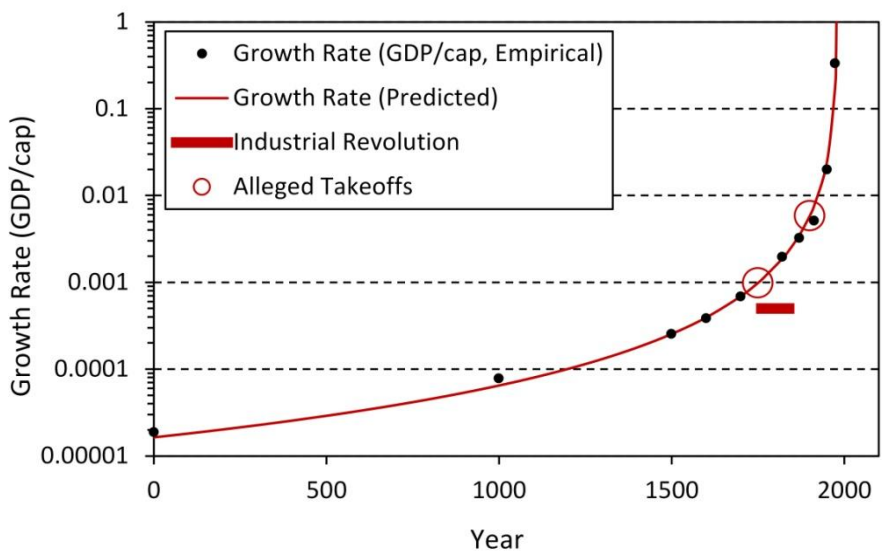

**Figure 6.** *Growth rate of income per capita (GDP/cap). Empirical growth rate calculated using Maddison's data (Maddison, 2001) and interpolated gradients is compared with the predicted growth rate calculated using parameters of the fitted hyperbolic distributions displayed in Figures 1 and 2. Galor's claims (Galor, 2005a, 2011) are remarkably contradicted by the analysis of Maddison's data (Maddison, 2001), precisely the same data, which he used but never properly analysed. Galor's mystery of "the sudden spurt" in the growth rate of income per capita (Galor, 2005a, p. 220) is solved – there was no sudden spurt.*

In order to support his preconceived ideas, Galor ignored not only the analysis carried out over 50 years ago (von Foerster, Mora & Amiot, 1960) but also research contributions of his many other predecessors (Biraben, 1980; Clark,1968; Cook,1960; Durand, 1967, 1974, 1977; Gallant, 1990; Haub, 1995; Hoerner (1975); Kapitza, 2006; Kremer, 1993; Livi-Bacci, 1997; McEvedy & Jones, 1978; Podlazov, 2002; Shklovskii, 1962, 2002; Taeuber & Taeuber, 1949; Thomlinson, 1975; Trager, 1994). Galor's claims are in conflict with science. They are not just unsupported by science – they are repeatedly contradicted by the scientific analysis of data and most notably by the analysis of precisely the same data, which he used during the formulation of his theory.

Mathematical analysis of Maddison's data (Maddison, 2001), precisely the same data as used by Galor, solves also his mystery "of the sudden spurt in growth rates of output per capita" (Galor, 2005a, p. 220) – there was no spurt. Results of analysis are presented in Figure 6. Growth rate of income per capita (GDP/cap) was increasing monotonically. Industrial Revolution did not accelerate the growth of income per capita. The postulated takeoffs from stagnation to growth (yet another mystery of growth claimed by Galor) did not happen because there was no stagnation and because the growth rate was increasing steadily without any major interruption. The only real mystery is why the growth rate of income per capita was so remarkably stable over such a long time.

Hyperbolic distributions, which increase monotonically, are characterised by monotonically-increasing growth rates, as shown in Figures 5-6. Claiming the existence of sudden spurts in such distributions is scientifically unjustifiable. Going a step further and claiming that such an imaginary spurt is a mystery, which needs to be explained encourages other researchers to carry out pointless and unproductive research.

It is useful to compare the mathematical analysis of Maddions's data presented in Figure 6 with the distorted presentation used by Galor reproduced in Figure 7. Both figures are based on *precisely the same set of data* (Maddison, 2001). The contrast is striking.

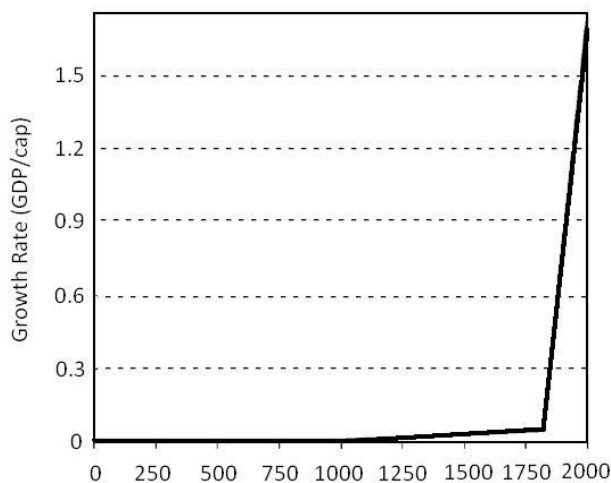

**Figure 7.** *Galor's distorted, strongly suggestive and misleading presentation of Maddison's data (Maddison, 2001) describing the growth rate of output (income) per capita (Galor, 2005a, p. 179). Precisely the same data, when correctly displayed and analysed (see Figure 6), show that "the sudden spurt in the growth rate of output per capita" claimed by Galor (2005a, p. 220) did not exist.*

While the data and their analysis, displayed in Figure 6, present monotonically-increasing growth rate of income per capita, Galor's distorted presentation of precisely the same data show a clear "sudden spurt." Maybe Galor was so strongly guided by the traditional interpretations of economic growth that he could not accept the clear contradicting evidence or maybe he simply did not know how to analyse data. In any case, intentional or unintentional, such ubiquitous distorted diagrams used repeatedly in his theory can be hardly expected to lead to reliable conclusions. On the contrary, they can be expected to lead only to incorrect conclusions.

Galor gives also many isolated examples of small growth rates in the past and significantly larger values at a later stage of growth but all these examples are not only meaningless but also strongly misleading. They reflect nothing more than just the natural properties of hyperbolic distributions. Using them to prove stagnation and transitions from stagnation to growth is scientifically irresponsible.

Of course growth rates of income per capita (GDP/cap) were small over a long time and significantly larger at a certain later stage of growth because they were following monotonically-increasing second-order hyperbolic distributions [see eqn (12)]. Hyperbolic distributions (second-order or first-order) are slow over a long time and fast over a short time but they are still the same, monotonically-increasing distributions. They are not characterised by sudden spurts. There is no profound mystery about them that needs to be explained by some elaborate research or mathematical formulations. It is just a simple and straightforward hyperbolic growth. The mystery is solved. Picking up some isolated numbers from such hyperbolic distributions and drawing some profound conclusions based on such examples is unscientific. The only mystery that needs to be explained is why the economic growth and the growth of population were hyperbolic and why they were so remarkably stable (undisturbed) over such a long time.

## Summary and conclusions

Galor discovered many "mysteries of the growth process" (Galor, 2005a, p. 220). One of his mysteries is "the sudden spurt in growth rates of output per capita and population that occurred in the course of the take-off from stagnation to growth"(Galor, 2005a, p. 220).

His discoveries are based on the crude and distorted presentations of data (Ashraf, 2009; Galor, 2005a, 2005b, 2007, 2008a, 2008b, 2008c, 2010, 2011, 2012a, 2012b, 2012c; Galor and Moav, 2002; Snowdon & Galor, 2008). His mysteries are of his own creation. They do not need to be explained because they do not exist. They describe the world of fiction.

Historical economic growth and the growth of human population were hyperbolic (Kapitza, 2006; Kremer, 1993; Nielsen, 2014, 2016a, 2016b, 2016c, 2016d, 2016e; Podlazov, 2002; Shklovskii, 1962, 2002; von Foerster, Mora & Amiot, 1960; von Hoerner, 1975). Hyperbolic distributions are monotonic and they are characterised by the monotonically-increasing growth rates. Sudden spurts do not exist in such distributions.

It is essential to understand that it is incorrect to take a slowly-increasing distribution and automatically claim the evidence of Malthusian stagnation. The state of stagnation might occur when there is a strong interference between a primary force propelling growth and some random opposing forces. Effects of stagnation should be reflected in the growth trajectory, which should be clearly unstable. If the growth follows a steadily-increasing trajectory without any clear signs of random behaviour then there is no need to complicate the description of the mechanism of growth by introducing random forces whose presence is undetectable. The fundamental principle of scientific investigation is to look for the simplest descriptions and explanations. Introducing unnecessary complications is simply unscientific.

It appears that the established knowledge in demography and in economic research is strongly based on a series of postulates revolving around the concept of Malthusian stagnation and around the alleged transition from the imagined stagnation to growth. Complicated mechanisms and interactions are then used (and gradually made even more complicated) to explain the mechanism of growth. Galor went one step further and reinforced these incorrect interpretations by his persistently distorted presentations of data (such as shown in Figure 7) and by his repeated quotations of certain well-selected figures to support his preconceived ideas, figures which were supposed to illustrate the concepts of stagnation and takeoffs from stagnation to growth but when closely analysed illustrate nothing else than the simple hyperbolic growth. All such complicated explanations are contradicted by data. Close examination of data shows that there was no stagnation and no transition from stagnation to growth. Data show consistently that the mechanism of the economic growth and of the growth of human population must have been simple because hyperbolic growth is exceptionally simple [see eqn (5)].

Some types of growth might be slow and stagnant but hyperbolic growth is not stagnant even when it is slow. It is prompted by the same mechanism during the time when it is slow and when it is fast. If the mechanism of Malthusian stagnation is used to explain the slow hyperbolic growth, precisely the same mechanism should be used to explain the fast growth, which is commonly described as explosion. It is incorrect to divide hyperbolic distributions into two or three sections and assign different mechanisms of growth to each of such arbitrarily selected sections. Hyperbolic distributions have to be explained as a whole and the same mechanism should be applied to the apparent slow and to the apparent fast sections.

It is incorrect to take a hyperbolic distribution and look for a sudden takeoff from the imagined stagnation to growth, as Galor did repeatedly. It is impossible to divide hyperbolic distribution into such distinctly different sections and the best way to see it, is by using the reciprocal values (Nielsen, 2014) because hyperbolic distribution is then represented by a decreasing straight line and it is obvious that it is impossible to claim a change of direction on the straight, which shows no change of direction.

Hyperbolic growth is not the only type of growth that can be slow over a certain time but not stagnant. Exponential growth is initially slow but it gradually becomes faster. At a certain stage, as if suddenly, it becomes overwhelmingly fast, the effect described as "the second half of the chessboard" (Kurzweil, 1999). Logistic growth is also initially slow but it is not stagnant.

The difference between hyperbolic and exponential distributions is that for hyperbolic distributions the apparent (but non-existent) transition from a slow to fast growth is more clearly articulated. That is why hyperbolic distributions are so often misinterpreted, particularly if they are distorted as it is done repeatedly and persistently in Galor's publications. However, this apparent transition from slow to fast growth does not happen at any given time or even over a certain specific range of time. It happens all the time. The acceleration is gradual along the entire range of hyperbolic distribution.

Growth of income per capita (GDP/cap) is represented by the ratio of two hyperbolic distributions (Nielsen, 2015). The ratio of monotonically increasing-distributions is characterised by the monotonically-increasing growth rate. We have shown that while the growth rate of the GDP and population increases hyperbolically, the growth rate of income per capita (GDP/cap) increases by following a second-order hyperbolic distribution (the reciprocal of the second order polynomial). There is no sudden spurt in any of these distributions and we have demonstrated this point by using the world economic growth and the growth of human population.

When doctrines accepted by faith are defended, scientific rules of engagement are readily violated. Contradicting data are then either ignored or manipulated to support preconceived ideas. Economic and demographic research has no place for this type of activities. However, intentionally or unintentionally, such unscientific approach to research appears to have been adopted during the formulation of the Unified Growth Theory (Galor, 2005a, 2011). Numerous preceding research works (e.g. Biraben, 1980; Clark,1968; Cook,1960; Durand,1967, 1974, 1977; Gallant, 1990; Haub, 1995; Kapitza, 2006; Kremer, 1993; Livi-Bacci, 1997; McEvedy & Jones, 1978; Podlazov, 2002; Shklovskii, 1962, 2002; Taeuber & Taeuber, 1949; Thomlinson, 1975; Trager, 1994; von Foerster, Mora and Amiot, 1960; von Hoerner, 1975) were ignored and the excellent data of Maddison (2001) were manipulated and distorted to support a series of preconceived ideas.

Hyperbolic distributions may be confusing but there is no excuse for distorting them to make them even more confusing. There is also no excuse for failing to analyse hyperbolic distributions because their analysis is trivially simple (Nielsen, 2014). The analysis of the growth rates is in the same category. Graphically, all these distributions become abundantly clear by using either the semilogarithmic scales of reference of by displaying the reciprocal values of growing entities or of their corresponding growth rates.

Galor's Unified Growth Theory is scientifically unacceptable and so are also many traditional interpretations of economic growth and of the growth of human population, interpretations based on the incorrect understanding of the mathematical properties of hyperbolic distributions. The recent and readily-accessible Maddison's data (Maddison, 2001, 2010) make it now easy to study the mechanism of the historical economic growth and of the growth of human population. They demonstrate that certain fundamental postulates revolving around the concept of Malthusian stagnation and around the assumed transition from the non-existent stagnation to growth, still used routinely in the established knowledge in demography and in economic research, are no longer acceptable.